# ИНФОРМАТИКА

УДК 004.9:66.013.512

## МОДУЛЬНАЯ ТЕХНОЛОГИЯ РАЗРАБОТКИ РАСШИРЕНИЙ САПР: ПРОФИЛИ НАРУЖНЫХ СЕТЕЙ ВОДОСНАБЖЕНИЯ И КАНАЛИЗАЦИИ

В.В.Мигунов, Р.Р.Кафиятуллов, И.Т.Сафин

*ЦЭСИ РТ при КМ РТ, г. Казань*

Аннотация

Модульная технология разработки проблемно-ориентированных расширений САПР применена к задаче проектирования профилей наружных сетей водоснабжения и канализации с реализацией в программной системе TechnoCAD GlassX. Выявлено единство состава профилей водопровода и канализации, разработана системная модель чертежей профилей сетей, включающая структурированное параметрическое представление (свойства объектов и их связи, общие установки и установки по умолчанию) и операции над ним, автоматизирующие проектирование.

Библ.2

Abstract

V.V. Migunov, R.R. Kafiyatullov, I.T. Safin. The modular technology of development of the CAD expansions: profiles of outside networks of water supply and water drain // Izvestiya of the Tula State University/ Ser. Mathematics. Mechanics. Informatics. Tula: TSU, 2003. V._. N _. P. __–__.

The modular technology of development of the problem-oriented CAD expansions is applied to a task of designing of profiles of outside networks of water supply and water drain with realization in program system TechnoCAD GlassX. The unity of structure of this profiles is revealed, the system model of the drawings of profiles of networks is developed including the structured parametric representation (properties of objects and their interdependence, general settings and default settings) and operations with it, which efficiently automate designing.

Bibl.2

Настоящая работа посвящена применению модульной технологии разработки проблемно-ориентированных расширений систем автоматизированного проектирования (САПР) реконструкции предприятия, общие положения которой изложены в [1]. Объект приложения технологии - автоматизация подготовки чертежей профилей наружных сетей водоснабжения и канализации (ПНС), выполненная в САПР TechnoCAD GlassX. Чертежи ПНС включаются в состав рабочих чертежей системы проектной документации для строительства согласно стандарту [2].

Анализ показывает, что чертежи профилей наружных сетей водопровода и канализации имеют большую степень сходства, и их подготовку удобно автоматизировать в рамках одного специализированного расширения САПР. И те, и другие профили изображают развертку сети по оси трубопровода и допускают единое представление в чертеже в виде совокупности следующих элементов:

- линии поверхности земли (проектная – тонкой сплошной линией, натурная – тонкой штриховой линией);
- уровень грунтовых вод – тонкой штрих-пунктирной линией;
- графические обозначения надземных объектов (пересекаемых автодорог, железных дорог, трамвайных путей, пешеходных дорожек, эстакад и т.д.), врезанные в линию проектной поверхности земли, и текстовую информацию о них;
- подземные инженерные сооружения и сети (сечения труб в виде эллипсов и кабелей в виде закрашенных эллипсов), влияющие на прокладку проектируемого трубопровода, с указанием их габаритных размеров и высотных отметок;
- изображение профиля проектируемого трубопровода;
- изображения колодцев и дождеприемников проектируемого трубопровода;
- изображения подземных частей зданий и сооружений, связанных с проектируемым трубопроводом;
- графические обозначения защитных футляров на трубопроводе с указанием диаметров, длин и привязок их к оси дорог или проектируемым сетям и сооружениям;
- размеры, отметки высоты и текстовые обозначения;
- таблицу основных данных для прокладки трубопровода;
- масштаб изображения профиля.

В этом списке выделяются изображения подземных частей зданий и сооружений, чертежи которых слабо связаны с трассой трубопровода и не допускают автоматической генерации по комплекту параметров ПНС. Они исключаются из автоматизируемой в этом расширении САПР части работ и создаются обычными графическими средствами САПР.

Параметрическое представление объектов и их связей в ПНС содержит следующие списки, в описании которых отражены принадлежности (ссылки). Кроме того, учтена необходимость специфицирования ПНС. Все координаты и размеры задаются в миллиметрах натурных размеров от единой базовой точки на чертеже, если не указано иначе.

*Поверхности земли и уровень грунтовых вод*

Задаются последовательностями двумерных точек с неубывающей X-координатой и представляют собой ломаные с типом линии GlassX сплошная тонкая, штриховая или штрих-пунктирная тонкая соответственно. Каждая ломаная может иметь свой цвет. Эти объекты не объединены в список и могут быть отнесены к общим установкам ПНС.

*Надземные объекты*

Список надземных объектов, графические обозначения которых автоматически врезаются в линию проектной поверхности земли, для каждого из объектов содержит его тип (автодорога, железная дорога, эстакада 1, эстакада 2), X-координату оси, текстовое обозначение (на чертеже размещается на оси непосредственно над таблицей основных данных) и цвет, а также, в зависимости от типа, высоту и (или) ширину.

*Сечения труб и кабелей*

Список сечений труб и кабелей для каждого сечения содержит тип (сечение трубы, сечение кабеля, сечение телефонной канализации), координаты оси и цвет. Для трубы дополнительно задаются диаметр, толщина стенки, а также могут задаваться текст условного обозначения (на чертеже размещается на оси непосредственно над таблицей основных данных), наличие футляра, диаметр, толщина стенки и длина футляра.

*Точки углов поворота*

Список точек углов поворота для каждой точки содержит X-координату, текст, размещаемый на чертеже на оси непосредственно над таблицей основных данных, и текстовое обозначение точки угла поворота (на чертеже размещается в строке "Номер колодца, точки угла поворота" таблицы основных данных).

*Колодцы и дождеприемники*

Список колодцев и дождеприемников для каждого из них содержит тип (колодец, дождеприемник), X-координату оси, диаметр (ширину), переход за низ трубы (используется для автоматического определения глубины колодца в зависимости от глубины заложения трубопровода), расстояние (в мм бумаги) от проектной поверхности земли до текстового обозначения глубины заложения трубопровода (на чертеже размещается над колодцем), текстовое обозначение (на чертеже размещается в строке "Номер колодца, точки угла поворота" таблицы основных данных), цвет и тип линии (сплошная основная или сплошная тонкая).

*Защитные футляры*

Список защитных футляров для каждого футляра содержит X-координату центра графического обозначения, тип связи диаметра футляра с диаметром трубы и параметр связи, толщину стенки, длину и цвет. Тип связи диаметра футляра с диаметром трубы и параметр связи используются для автоматического изменения диаметра футляра при изменении диаметра трубы. Имеется два варианта:
- диаметр футляра изменяется пропорционально диаметру трубы, т.е. больше диаметра трубы в фиксированное число раз (параметр связи);
- диаметр футляра больше диаметра трубы на фиксированную величину (параметр связи).

*Типы труб*

Список типов труб для каждого типа содержит наружный диаметр (используется для изображения труб на чертеже), наименование, материал и тип изоляции труб (на чертеже размещаются в строке "Обозначение трубы и тип изоляции" таблицы основных данных), а также специфицирующие свойства, используемые для автоматической генерации спецификаций: позиционное обозначение; обозначение; масса единицы, кг; примечание; тип, марка, обозначение документа, опросного листа; наименование и техническая характеристика; единица измерения; наименование завода-изготовителя; код оборудования, изделия, материала.

*Трубы*

В списке труб, составляющих профиль проектируемого трубопровода, под трубой понимается набор продолжающих друг друга труб одного типа, идущих под разными уклонами. Каждая такая труба имеет свой тип (ссылку на список типов труб) и цвет и задается последовательностью двумерных точек с неубывающей X-координатой, образующей ломаную, являющуюся осью трубы.

*Тексты*

Список текстов для каждого из них содержит собственно многострочный текст, установку шрифта, шаг строк, цвет, координаты точки начала текста. От одного текста могут идти одна или более сносок к сечениям труб (с футлярами), колодцам или дождеприемникам, защитным футлярам. Для указания диаметров футляров используется специальный символ "Диаметр".

*Сноски от текстов к сечениям труб (с футлярами)*

Список сносок от текстов к сечениям труб (с футлярами) для каждой сноски содержит ссылки на текст из списка текстов и на сечение трубы (с футляром) из списка сечений труб и кабелей, а также двумерный вектор смещения (в мм бумаги) точки указания сноски от оси трубы. Цвет сноски берется из цвета ее текста.

*Сноски от текстов к защитным футлярам*

Список сносок от текстов к защитным футлярам для каждой сноски содержит ссылки на текст из списка текстов и на защитный футляр из списка защитных футляров, а также двумерный вектор смещения (в мм бумаги) точки указания сноски от центра графического обозначения футляра. Цвет сноски берется из цвета ее текста.

*Сноски от текстов к колодцам и дождеприемникам*

Список сносок от текстов к колодцам и дождеприемникам для каждой сноски содержит ссылки на текст из списка текстов и на колодец или дождеприемник из списка колодцев и дождеприемников, а также двумерный вектор смещения (в мм бумаги) точки указания сноски от середины низа колодца или дождеприемника. Цвет сноски берется из цвета ее текста.

*Размеры*

Все размеры являются горизонтальными цепными размерами, при этом обычный линейный размер – частный случай цепного. Список размеров для каждого из них содержит ссылки на надземные объекты, сечения труб и кабелей, точки углов поворота, колодцы и дождеприемники, защитные футляры. Все перечисленные объекты имеют оси, проводимые на чертеже до верха таблицы основных данных и выступающие в качестве выносных линий размера. Также для каждого размера хранятся смещения (в мм бумаги) вдоль этих осей размерной линии от верха таблицы и текстов размера от размерной линии.

Индивидуальных установок (наличие сноски и флаг проведения ее к концу полки, длина засечки, установка шрифта и цвет) размеры не имеют, они подчиняются общим для профиля установкам размеров. В тексты размеров автоматически проставляются имеющиеся реальные значения этих размеров в метрах с двумя цифрами после десятичной точки (по ГОСТ 21.604-82).

*Отметки высоты*

Список отметок высоты для каждой отметки содержит ссылку на сечение из списка сечений труб и кабелей, смещение точки указания стрелкой вправо вдоль выносной линии (положительное или отрицательное), направление полки (влево или вправо), смещение полки вверх от выносной (положительное или отрицательное). Своего цвета отметки высоты не имеют, применяется общая для профиля установка цвета.

В приведенных списках возникают связи принадлежности у объектов, отвечающих за оформление чертежа ПНС (сноски привязаны к текстам, защитным футлярам, колодцам и дождеприемникам; размеры – к надземным объектам, сечениям труб и кабелей, точкам углов поворота, колодцам и дождеприемникам, защитным футлярам; отметки высоты – к сечениям труб и кабелей). Эти связи позволяют автоматизировать различные операции по модификации ПНС. Например, при удалении колодца автоматически удаляются связанные с ним сноски текстов и выносные линии размеров (сами размеры автоматически перегенерируются или удаляются). При переносе сечения трубы или кабеля автоматически перегенерируются связанные с сечением сноски, а также переносятся связанные с сечением выносные линии размеров и отметки высоты (сами размеры и тексты отметок высоты автоматически перегенерируются).

Кроме связей принадлежности, также возникают и координатные связи. Так, глубина колодца вычисляется автоматически в зависимости от следующих параметров: проектная отметка земли и отметка низа трубы в месте расположения колодца; переход колодца за низ трубы. При этом изображение входа трубы в колодец вычерчивается автоматически. Надземные объекты, исходя из X-координаты их оси, автоматически врезаются в проектную поверхность земли.

*Установки профиля наружной сети*

Установки ПНС делятся на установки, действующие по умолчанию, и общие для всего профиля установки. Установки, действующие по умолчанию, используются при добавлении объектов в ПНС.

Общие для всего профиля установки:
- установки таблицы основных данных: координаты правого верхнего угла шапки таблицы; наличие шапки у таблицы и минимальная длина таблицы без шапки (используются для продолжения таблицы вправо и стыковки с таблицей основных данных другого профиля); установка шрифта для текстов таблицы и для обозначения масштабов по горизонтали и по вертикали; единица измерения уклонов в таблице (промилле или проценты);
- установки вспомогательной шкалы: наличие, цена деления, цвет шкалы, установка шрифта для обозначений высоты на делениях шкалы;
- глобальные установки построения: масштабы по горизонтали и по вертикали (по ГОСТ 21.604-82), тип трубопровода (водопровод, канализация – от типа трубопровода зависит отрисовка труб внутри колодцев), основание (грунт) – выводится в соответствующей строке таблицы основных данных, минимальное отношение длин главных осей эллипсов (если эллипс слишком вытягивается, он рисуется потолще с учетом этого ограничения), условный уровень проектной поверхности земли (используется для определения верха колодцев и положения надземных объектов пока нет реального уровня, а также как верхнее предельное значение вспомогательной шкалы и центр обозначения масштабов), условный уровень низа проектируемого трубопровода (используется для определения низа колодцев, вертикального положения футляров и позиционирования отметок высоты на сечениях труб пока нет реального уровня), цвета линий проектной и натурной поверхностей земли и уровня грунтовых вод, установка шрифта для нанесения глубины заложения трубопровода, обозначений сечений труб и надземных объектов;
- для сечений труб и кабелей: диаметр (ширина) изображений сечений кабелей (соответствует диаметру круга при совпадении масштабов по горизонтали и по вертикали и является минимальным из габаритов эллипса при различии этих масштабов), длины условных обозначений сечений труб, кабелей и телефонной канализации, располагаемых на оси непосредственно над таблицей основных данных, длина (продольный катет) и размах крыльев (двойной поперечный катет) стрелки условных обозначений сечений кабелей, диаметр закрашенного круга условных обозначений сечений телефонной канализации;
- цвет точек углов поворота;
- условный диаметр трубы, используемый для определения диаметров футляров, расположенных не на трубе (например, во время добавления футляра, когда курсор не попадает ни на одну из имеющихся труб);
- для размеров: наличие сноски и флаг проведения ее к концу полки, длина засечки, установка шрифта и цвет;
- для отметок высоты: тип линии (сплошная основная или сплошная тонкая) и длина катета стрелки, установка шрифта и цвет.

Отметим, что для ПНС задаются два масштаба: по горизонтали (1:500 – 1:1500) и по вертикали (1:100 – 1:500). При этом все координаты и размеры запрашиваются и хранятся во внутреннем представлении в реальной системе координат (Натура), а автоматическая генерация чертежа выполняется в соответствии с заданными масштабами, что значительно облегчает работу проектировщика.

Установки, действующие по умолчанию:
- тип (автодорога, железная дорога, эстакада 1, эстакада 2) и цвет для надземных объектов;
- тип (сечение трубы, сечение кабеля, сечение телефонной канализации), диаметр и толщина стенки труб, наличие защитного футляра на трубах и цвет для сечений труб и кабелей;
- тип (колодец, дождеприемник), диаметр (ширина) и переход за низ трубы, используется для автоматического определения глубины колодца в зависимости от глубины заложения трубопровода), расстояние (в мм бумаги) от проектной поверхности земли до текстового обозначения глубины заложения трубопровода (на чертеже размещается над колодцем), цвет и тип линии (сплошная основная или сплошная тонкая) для колодцев и дождеприемников;
- тип связи диаметра футляра с диаметром трубы и параметр связи (используются для автоматического изменения диаметра футляра при изменении диаметра трубы; тип связи имеет два варианта: диаметр футляра изменяется пропорционально диаметру трубы, т.е. больше диаметра трубы в фиксированное число раз (параметр связи); диаметр футляра больше диаметра трубы на фиксированную величину (параметр связи)), толщина стенки и длина (в мм бумаги), цвет для защитных футляров;
- цвет и последний использованный тип для труб;
- установка шрифта, шаг строк, цвет, положение начала сноски, наличие второй полки при двух строках для текстов;
- смещение текстов размера от размерной линии вдоль выносных;
- смещение точки указания стрелкой вправо вдоль выносной линии (положительное или отрицательное), направление полки (влево или вправо), смещение полки вверх от выносной (положительное или отрицательное) для отметок высоты.

Описанное параметрическое представление ПНС может сохраняться на диске как комплект параметров, без геометрической части. Однако при выборе комплекта параметров на диске для работы с ним пользователь просматривает изображение ПНС, которое генерируется при перемещении по меню в режиме on-line, как это показано на рис.1. Отметим компактность хранения комплекта параметров - для изображенного на рис.1 ПНС он занимает на диске 474 байта.

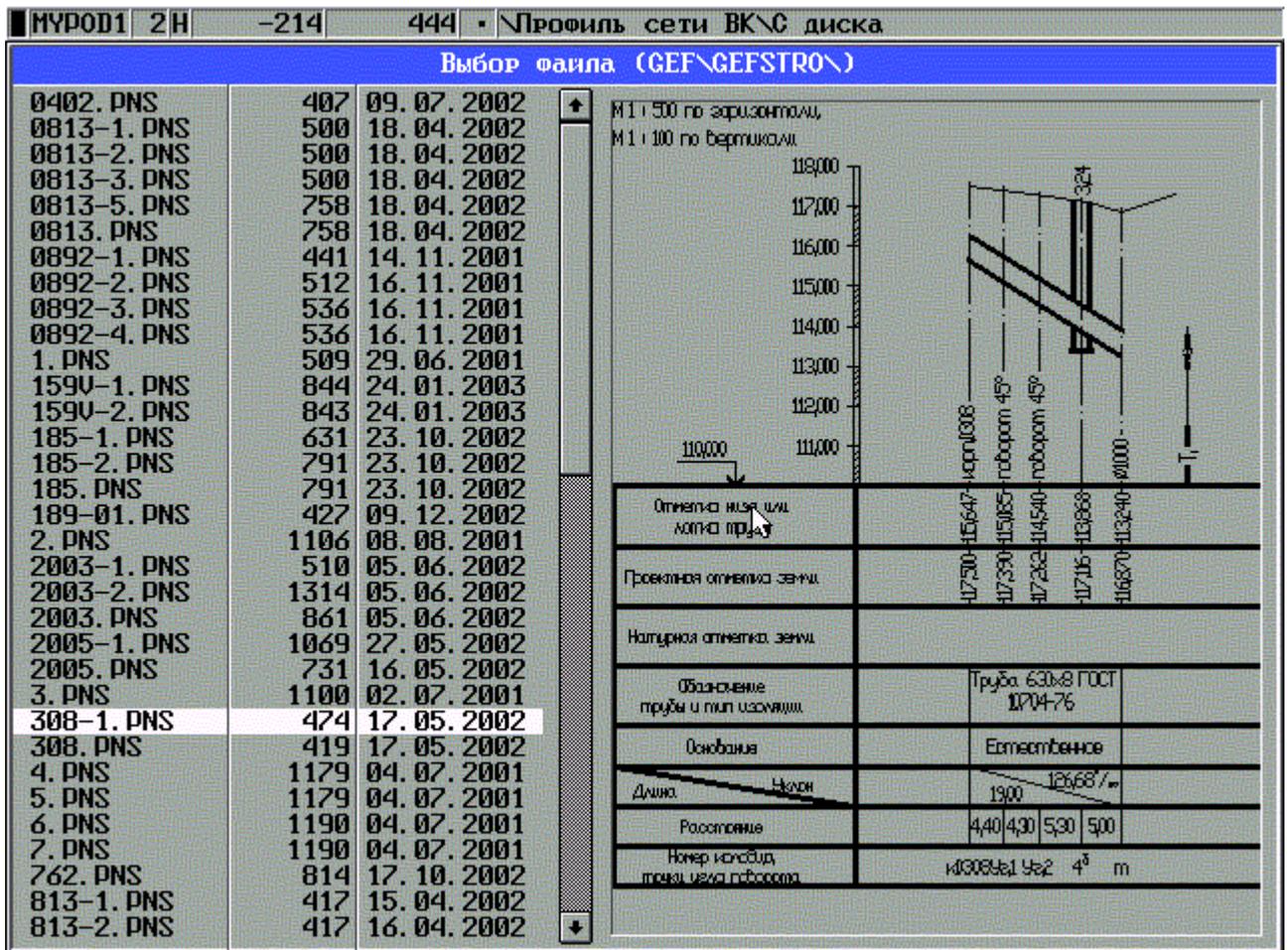

Рис.1. Выбор прототипа из ранее сохраненных на диске комплектов параметров ПНС

*Операции по подготовке чертежа ПНС*

Над параметрическим представлением ПНС проектировщик последовательно выполняет различные операции, выбирая их из основного меню.

"С диска" – чтение из дискового файла ранее сохраненного в нем внутреннего представления ПНС.

"Из чертежа" – взятие параметров ПНС из выбираемого в чертеже профиля.

"Добавление" – группа операций по добавлению в профиль надземных объектов, сечений труб и кабелей, точек углов поворота, колодцев и дождеприемников, защитных футляров, труб, текстов, сносок, размеров, отметок высоты, поверхностей земли и уровня грунтовых вод. Здесь же осуществляется добавление выносных линий в цепные размеры, а также продолжение ломаных поверхностей земли и уровня грунтовых вод.

"Удаление" – запускается режим повторяющегося удаления объектов ПНС (надземных объектов, сечений труб и кабелей, точек углов поворота, колодцев и дождеприемников, защитных футляров, труб, текстов, сносок, размеров, отметок высоты, поверхностей земли и уровня грунтовых вод), выбираемых курсором по одному.

"Перенос" – запускается циклический режим переноса объектов ПНС (надземных объектов, сечений труб и кабелей, точек углов поворота, колодцев и дождеприемников, защитных футляров, текстов, сносок текстов, размеров, отметок высоты, вершин ломаных поверхностей земли и уровня грунтовых вод, стыков и концов труб), выбираемых курсором по одному.

"Редактирование" – группа операций по редактированию объектов ПНС, см. ниже.

"Копирование" – запускается циклический режим копирования объектов ПНС (надземных объектов, сечений труб и кабелей, колодцев и дождеприемников, защитных футляров, текстов, отметок высоты), выбираемых курсором по одному.

"Свойства" – запускается циклический режим изменения свойств объектов ПНС (надземных объектов, сечений труб и кабелей, точек углов поворота, колодцев и дождеприемников, защитных футляров, труб, текстов, размеров, отметок высоты), выбираемых курсором по одному.

"Перенос профиля" – перенос всего профиля в другое место на поле чертежа.

"Установки" – изменение установок, а также модификация списка типов труб.

"На диск" – запись на диск текущего параметрического представления профиля. Его можно повторно использовать в дальнейшем, копировать на другие рабочие места.

"В чертеж" – помещение профиля в чертеж.

*Редактирование объектов профиля*

Здесь реализуются следующие операции:

"Продолжение трубы" – предлагается выбрать конец трубы для продолжения среди подсвеченных крестиками. После этого последовательно задаются новые стыки и конец трубы сначала указанием в чертеже (с подсветкой строящегося трубопровода), а затем уточнением координат в форме ввода (с возможностью указания в чертеже уже имеющихся концов и стыков труб, подсвеченных крестиками).

"Разрезание участка трубы" – сначала в чертеже выбирается труба. По X-координате курсора определяется новый стык трубы. Затем предлагается задать положение нового стыка (сначала указанием в чертеже с подсветкой, а затем в форме ввода уточнения координат).

"Разрезание отрезка проектной поверхности земли", "Разрезание отрезка натурной поверхности земли", "Разрезание отрезка уровня грунтовых вод" – опции доступны, если поверхности земли и уровень грунтовых вод (соответственно) уже нанесены. Сначала курсором на поверхности земли или уровне грунтовых вод указывается положение новой вершины (по X-координате курсора), после чего предлагается перенести новую вершину (указанием в чертеже с подсветкой).

"Удаление стыка или конца трубы" – сначала в чертеже выбирается труба, затем выбирается удаляемый стык или конец трубы (подсвечены крестиками). Если труба состояла только из одного участка, то она удаляется целиком.

"Удаление вершины проектной поверхности земли", "Удаление вершины натурной поверхности земли", "Удаление вершины уровня грунтовых вод" – опции доступны, если поверхности земли и уровень грунтовых вод (соответственно) уже нанесены. Среди подсвеченных крестиками вершин предлагается выбрать удаляемую. Если ломаная состояла только из одного отрезка, то она удаляется целиком.

"Деление трубы на две" – в чертеже выбирается труба, затем один из ее стыков, подсвеченных крестиками, по которому произойдет деление трубы на две.

"Слияние труб" – в чертеже выбирается один из подсвеченных крестиками совпадающих концов труб, по которому произойдет объединение двух соседних труб с указанным совпадающим концом. При этом, если тип или цвет труб были различны, предлагается задать их в форме ввода.

"Редактирование текста" – в чертеже выбирается текст, после чего его можно отредактировать.

"Правка данных в таблице" – изменение в таблице основных данных числовых значений в строках "Проектная отметка земли", "Натурная отметка земли", "Длина\Уклон", "Расстояние":

Для строк "Проектная отметка земли" и "Натурная отметка земли" после задания в форме ввода нового значения отметки земли предлагается выбрать один из следующих вариантов изменения ломаной поверхности земли: добавить вершину, сдвинуть левый или правый конец отрезка по вертикали, сдвинуть отрезок по вертикали.

Для строки "Длина" после задания в форме ввода нового значения горизонтальной составляющей длины участка трубы предлагается выбрать один из следующих вариантов изменения горизонтальной составляющей длины участка трубы: сдвинуть по горизонтали находящиеся слева или справа трубы (изменяя уклон участка), сдвинуть расположенные слева или справа трубы, не изменяя уклона участка,

Для строки "Уклон" после задания в форме ввода нового значения уклона участка трубы предлагается сдвинуть по вертикали находящиеся слева или справа трубы.

Для строки "Расстояние" после задания в форме ввода нового значения расстояния предлагается сдвинуть колодцы и точки угла поворота, находящиеся слева или справа.

Таким образом, изменения в ПНС можно вносить, работая с изображением профиля, или внося изменения в таблицу – эти две части синхронизируются автоматически.

*Таблица основных данных*

Изображение таблицы основных данных генерируется полностью автоматически на основе текущего состояния ПНС. Автоматическое вычерчивание таблицы происходит следующим образом:

Высота строк таблицы и ширина графы с заголовками строк (шапки) задана ГОСТ 21.604-82, тип линий – сплошная основная, установки шрифта единые для всей таблицы.

Строка "Отметка низа или лотка трубы" заполняется при наличии труб, составляющих профиль проектируемого трубопровода. Отметки проставляются в точках углов поворота, вдоль осей колодцев и дождеприемников, а также в местах стыка труб.

Строки "Проектная отметка земли" и "Натурная отметка земли" заполняются при наличии проектной и натурной поверхностей земли соответственно. Отметки проставляются в точках углов поворота и вдоль осей колодцев и дождеприемников.

Строка "Обозначение трубы и тип изоляции" заполняется при наличии труб, составляющих профиль проектируемого трубопровода. Если трубы имеют разные типы, то в эту строку автоматически заносится несколько обозначений.

Строка "Основание" может содержать только одно значение для всего профиля, задаваемое проектировщиком.

Строка "Длина\Уклон" заполняется при наличии труб, составляющих профиль проектируемого трубопровода. В строку автоматически заносятся длины и уклоны участков трубопровода с постоянным уклоном.

Строки "Расстояние" и "Номер колодца, точки угла поворота" заполняется при наличии точек углов поворота, колодцев и дождеприемников. В строку "Расстояние" автоматически заносятся расстояния между соседними объектами, а в строку "Номер колодца, точки угла поворота" их текстовые обозначения.

*Выводы*

Изложенные в настоящей работе состав и особенности реализации операций подготовки ПНС вместе с их структурированным параметрическим представлением являются системной моделью чертежей профилей наружных сетей водоснабжения и канализации, используемой при проектировании реконструкции предприятий согласно системе проектной документации для строительства. Модель первоначально разработана в 2001 году по модульной технологии [1] и в дальнейшем проявила высокую степень устойчивости к добавлению новых возможностей. За два года в рамках тех же модельных представлений в подготовку ПНС были добавлены возможности специфицирования (рис.2)

Рис.2. Типы труб: выбор из электронных каталогов

Дополнения вносились практически без изменения уже имевшихся списков объектов и операций. Это говорит как о высокой степени адекватности параметрического представления и списка операций объективной сущности ПНС, так и об эффективности модульной технологии разработки проблемно-ориентированных расширений САПР реконструкции предприятия.

## Литература